\documentclass[12pt,a4paper,final]{iopart}

\newcommand{\degree}{\ensuremath{^\circ}}

\usepackage{iopams}  
\usepackage{graphicx}
\usepackage[breaklinks=true,colorlinks=true,linkcolor=blue,urlcolor=blue,citecolor=blue]{hyperref}

\begin{document}

\title{Electron and phonon transport in twisted graphene nanoribbons}

\author{Aleandro Antidormi}
\address{Institut de Ci\`encia de Materials de Barcelona (ICMAB--CSIC)
             Campus de Bellaterra, 08193 Bellaterra, Barcelona, Spain}
\address{Dipartimento di Fisica, Universit\` degli Studi di Cagliari,
Cittadella Universitaria, I-09042 Monserrato (Ca), Italy}
\ead{aleandro.antidormi@dsf.unica.it}

\author{Miquel Royo}
\address{Institut de Ci\`encia de Materials de Barcelona (ICMAB--CSIC)
             Campus de Bellaterra, 08193 Bellaterra, Barcelona, Spain}
\ead{mroyo@icmab.es}
             
\author{Riccardo Rurali}
\address{Institut de Ci\`encia de Materials de Barcelona (ICMAB--CSIC)
             Campus de Bellaterra, 08193 Bellaterra, Barcelona, Spain}
\eads{rrurali@icmab.es, \mailto{rrurali@icmab.es}}

\vspace{10pt}
\begin{indented}
\item[]March 2017
\end{indented}

\begin{abstract}
We theoretically study the electrical, thermal and thermoelectric transport properties of 
graphene nanoribbons under torsional deformations. The modelling follows a nonequilibrium 
Green's function approach in the ballistic transport regime, describing the electrical and 
phononic properties through \textit{ab-initio} density functional theory and empirical
interatomic potentials, respectively. We consider two different types of deformations, 
a continuous twist of a given angle applied to the nanoribbon, and two consecutive twists 
applied in opposite angular directions. The numerical results are carefully analysed in
terms of spatially-resolved electron eigenchannels, polarization-dependent phonon transmission 
and thermoelectric figure-of-merit.

\end{abstract}

\vspace{2pc}
\submitto{\JPD}

\section{Introduction}

The field of research in two-dimensional (2D) materials has been enjoying
extraordinary growth during the past decade. This activity was triggered
by pioneering works on graphene~\cite{novoselov2004electric,
novoselov2005two,zhang2005experimental}, a 2D semimetallic allotrope of
carbon that turned out to
be an exceptionally fertile ground for advancing frontiers of condensed
matter physics~\cite{geim2007rise,neto2009electronic,novoselov2011nobel,
geim2011nobel}. Today the family of 2D materials has become densely populated with manifold specimens~\cite{ButlerACSNano13,BhimanapatiACSNano15,NovoselovScience16}: 
it comprises monolayer materials made of a single element (phosphorene, borophene, germanene and silicene) and
others featuring different atoms alternating in the same layer (boron nitride, transition metal
dichalcogenides (TMDCs) and Mxenes). 

Graphene consists of a hexagonal monolayer network of $sp^2$-hybridized
carbon atoms whose properties were expected to be outstanding, based on
theoretical predictions.
Unique ballistic transport properties, very long mean free path at room
temperature~ \cite{gunlycke2007room}, distinctive
integral and half-integral quantum hall effect\cite{novoselov2007room,
zhang2005experimental}, the highest known thermal
conductivity~\cite{balandin2011thermal} and high electron
mobility~\cite{du2008approaching} are among the most intriguing features of
graphene.
In particular, thanks to its electron mobility significantly higher than
that of the
widely-used Si, graphene has been envisioned as a candidate material for
new generations of nano-electronic devices~\cite{bonaccorso2010graphene,
bonaccorso2015graphene}.

From the experimental viewpoint, recent developments in fabrication
techniques have made
it possible to grow very narrow~\cite{li2008chemically,jiao2009narrow} and
atomically precise~\cite{cai2010atomically} graphene nanoribbons (GNRs),
further stimulating interest in this 2D material.
Nevertheless, like in any other real material, structural defects do exist
and can dramatically alter the properties of graphene.
Several experimental studies have shown, for instance, the occurrence of
intrinsic or extrinsic  defects
in graphene~\cite{hashimoto2004direct,gass2008free,meyer2008direct,
girit2009graphene,warner2009structural}. Others demonstrated how the
surface of fabricated graphene nanoribbons was not perfectly
flat due to the presence of {\em ripples}, i.e. nanoscopic roughening 
of the 2D plane~\cite{meyer2007structure,brivio2011ripples}.

These lattice imperfections have a strong influence on the electronic,
optical, thermal, and mechanical properties of the material.
As a matter of fact, many of the characteristics of technologically 
important materials such as the electrical conductance of semiconductors 
are governed by defects~\cite{kittel1966introduction}.
For this reason, defects are often deliberately introduced,
by irradiation or chemical methods. This is the basis for the 
shaping of the material properties via engineering of its defects, 
a possibility which determined 
the need  for a careful analysis of the relationship between
the nature and amount of graphene defects and the deriving physical
properties.

Several studies have, indeed, been devoted to the effect of structural
defects on graphene properties, with major attention given to point
defects, multiple vacancies and substitutional
impurities~\cite{banhart2010structural, liu2015defects,
haskins2011control}.
Furthermore, it has been proved that the occurrence of ripples in the surface
constitutes an intrinsic feature of 2D materials and its link with the
mechanical stability of the layer has been
explained~\cite{fasolino2007intrinsic,geim2007rise,meyer2007structure}.
For the specific case of GNRs, researchers have investigated the changes
in thermal and electron transport due to edge terminations, shape, width
and roughness~\cite{cresti2008charge,sevinccli2010enhanced,
li2011efficient,li2010phonon,areshkin2007ballistic}.
Recently, interest is growing also on the their mechanical and electronic
properties under different constraints such as mechanical
stress~\cite{bellido2012graphene,cadelano2010interplay, koskinen2012graphene,sanders2013theory} 
From an application point of view, 
their flexibility could allow their employment in stretchable electronics and in 
the creation of new-generation devices for non-linear energy harvesting~\cite{lopez2011nanostructured,lopez2013buckling,
lopez2014piezoelectric}. In this regard, of particular interest is twisting,
as this is a mechanical deformation unique from GNRs due to their exceptional
flexibility. Twisted nanoribbons have been fabricated encapsulated inside 
graphene nanotubes~\cite{KhlobystovACSNANO11, ChamberlainACSNANO12} or by means of 
chemical etching.~\cite{EliasNL10}
The effect of twisting on their structural,~\cite{Cranford2011,Nikiforov2014} electronic~\cite{Sadrzadeh2011} and 
transport~\cite{Al-Aqtash2013,Jia2014,Xu2015,Tang2012} properties
has been recently explored mostly from the theory side. Apart from showing the
expectable degradation of the electronic and thermal transport properties, 
calculations performed at different twist angles have predicted an electromechanical 
switching effect induced by twisting and occurring in both armchair~\cite{Jia2014,Xu2015} and zig-zag GNRs.~\cite{Al-Aqtash2013} 

In this work we investigate the electron and phonon transport properties of armchair
graphene nanoribbons by considering two alternative torsional deformation processes: 1) a fixed twist
of $180 \degree$ is applied along the longitudinal axis while the nanoribbon 
length is varied; 2) a twist of variable angle $\phi$ is applied for half the length of the
nanoribbon followed by a twist of $-\phi$ over the remaining half.
We study the evolution of the electrical and thermal conductance and analyse the 
results in terms of twist-induced structural deformation, spatially-resolved electron eigenchannels,
polarization-dependent phonon transport and thermoelectric properties. 

\section{Computational methods}
\label{sec:method}

\subsection{Electronic structure and electron transport}
\label{sub:elec-meth}

The electronic structure is calculated from first-principles,
using the implementation of density-functional theory (DFT) of
the {\sc Siesta} package~\cite{SolerJPCM02}. We account for
the core electrons with norm-conserving pseudopotentials and
use the Generalized Gradient Approximation (GGA) for the
exchange-correlation energy. We have optimized a light,
but accurate single-$\zeta$ polarized basis set to expand
the one-electron wavefunction that allowed us dealing efficiently
with relatively large systems.

We study an armchair graphene nanoribbon (GNR) of width 1.2~nm
and variable length. The edge dangling bonds are passivated
with hydrogen atoms. We sample the Brillouin zone with the
$\Gamma$ point only, because of the very large size of the
computational cell along the only periodic direction, taken to be
parallel to the $z$-axis. We use a $1 \times 1 \times 4$ {\bf k}-point 
mesh for the bulk calculations of the 3-unit cells structures 
required to compute the leads self-energies (see below).
All the structures were relaxed through a standard conjugate 
gradient algorithm until all the forces were smaller than 
0.04~eV/\AA.

\begin{figure}[t]
\centering
\includegraphics[width=0.8\columnwidth]{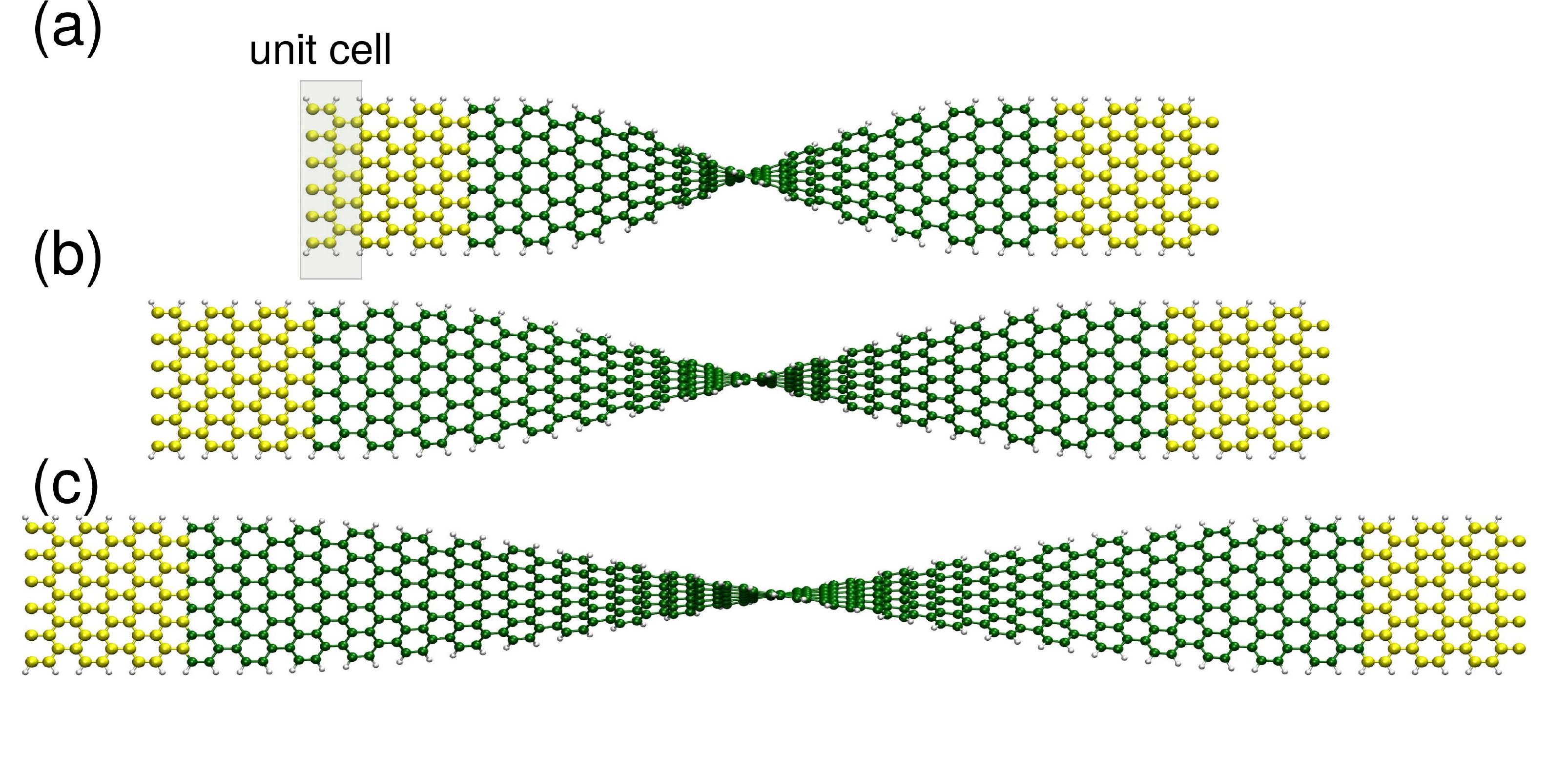}
\caption{Twisted armchair GNRs with a twist angle of 180\degree\ that
         develops over a length $L_{twist}$ of (a)~11, (b)~16, and 
         (c)~22 unit cells. Green and white spheres represent carbon 
         and hydrogen atoms, respectively. The two three-unit cell 
         regions on the left and on the right of the scattering region (shown as
         yellow spheres) are the leads and are kept frozen during the 
         relaxation and later modelled through the self-energies.}
\label{fig:ri}
\end{figure}

In the transport calculations we partition the 
system into three regions: a left lead, a right lead and a central 
scattering region that contains the twisted section of the ribbon. 
We solve the electronic transport problem in the central region 
with the {\sc transiesta} method~\cite{BrandbygePRB02} within the 
nonequilibrium Green's function formalism. The open boundary 
conditions imposed by the electrodes are accounted for through 
the left (right) self-energy $\Sigma_{L,R}(E)$.
The zero-bias transmission $T(E)$ is calculated as

\begin{equation}
 \mathcal{T}^e(E)
 = Tr \left[  \boldsymbol{\Gamma}_L(E) \, \mathbf{G}_C^r(E) \,
 \boldsymbol{\Gamma}_R(E)\, \mathbf{G}_C^a(E)  \right],
\label{eq:trans}
\end{equation}

\noindent where $\mathbf{G}_C^{r,a}(E)= [E\mathbf{S}_C-\mathbf{H}_C-\mathbf{\Sigma}_L^{r,a}(E)-\mathbf{\Sigma}_R^{r,a}(E)]$ 
is the retarded (advanced) Green's function of the scattering region, 
$\mathbf{\Gamma}_{L,R}(E)=i(\mathbf{\Sigma}_{L,R}^r - \mathbf{\Sigma}_{L,R}^a)$, 
$\mathbf{H}_C$ is the Hamiltonian matrix and $\mathbf{S}_C$ is the overlap matrix.
We use three unit cells of the armchair GNR for both electrodes (see Fig.~\ref{fig:ri} (a)), 
while the extent of the scattering region varies and depends 
on the torsional deformation considered in each case.
The electrical conductance $G^e(\mu)$ is then calculated through the Landauer formula as
\begin{equation}
G^e(\mu) = \frac{2e^2}{h} \int \mathcal{T}^e(E) \left(-\frac{\partial
f_0(E,\mu)}{\partial E} \right) dE,
\label{eq:land-e}
\end{equation}

\noindent where $f_0(E,\mu)$ is the Fermi-Dirac distribution function at a chemical potential $\mu$.

\subsection{Phonon transport}
\label{sub:phon-meth}

We simulate phonon transport with nonequilibrium 
Green's function methodologies on the same lead/scattering region/lead structures 
used for the electron transport calculations. For phonons, however, 
the use of DFT would be impractical due to the large number of calculations needed
to compute the force-constants matrix. Instead, we relax the structures and 
calculate the forces using the classical bond-order potential due to Brenner~\cite{BrennerPRB90} 
as implemented in the GULP code.~\cite{GaleMS03}

The phonon transmission function, $\mathcal{T}^{ph}(\omega)$, is obtained from the phonon equivalent
of Eq.~\ref{eq:trans}, where now the retarded (advanced) Green's function 
of the scattering region reads $\mathbf{G}_C^{r,a}(E)= [\omega^2 \mathbf{I}-\mathbf{F}_C-\mathbf{\Sigma}_L^{r,a}(\omega)-\mathbf{\Sigma}_R^{r,a}(\omega)]$, 
with $\mathbf{F}_C$ being the force-constant matrix, $\mathbf{I}$ the identity matrix and $\mathbf{\Sigma}_{L,R}(\omega)$
the self-energy of the left (right) contact.
In analogy with Eq.~\ref{eq:land-e}, the phononic thermal conductance 
is calculated within Landauer theory as
\begin{equation}
G^{ph}(T) = \frac{\hbar}{2\pi} \int \omega \mathcal{T}^{ph}(\omega)
\left(\frac{\partial n_0(\omega,T)}{\partial T} \right) d\omega,
\label{eq:land-ph}
\end{equation}
where $n_0$ is the equilibrium Bose-Einstein distribution function.

Besides, we employ a recently proposed~\cite{Ong2015} extension of the nonequilibrium Green's function method to 
disentangle the contribution of each individual phonon mode to the transmission function and the thermal conductance.
To this end we calculate the Bloch matrices of the leads as explained in Ref.~\cite{Ong2015} whose eigenvalues and
eigenvectors provide access, on the one hand, to the dispersion relations and eigendisplacements 
of the phonons at both leads and, on the other hand, to a single-mode transmission matrix, $\mathbf{t}(\omega)$. The square modulus of an element 
of this transmission matrix, $|t_{m,n}(\omega)|^2$, represents the probability of transmission from the $m$th phonon mode in the 
right lead to the $n$th mode in the left lead at a given frequency and has a value between 0 and 1. The sum of all single mode transmissions 
equals the result given by Eq.~\ref{eq:trans}.

Our transport simulations assume a ballistic regime, i.e., they are robust as long as electron-electron, phonon-phonon 
and electron-phonon interactions can be neglected. Despite such limitations, this a reasonably clean approach
where the only source of scattering is the structural deformation of the ribbon.

\subsection{Thermoelectric properties}
\label{sub:thermoelec-meth}

The thermoelectric figure of merit can be defined as,

\begin{equation}
 ZT=\frac{S^2 G^e T}{G^{ph}+\kappa^e},
 \label{eq_ZT}
\end{equation}

\noindent where $S$ is the Seebeck coefficient, $G^e$ is the electronic 
conductance, $T$ is the temperature, and $G^{ph}$ and $\kappa^e$ are 
the thermal conductance due to phonons and electrons, respectively. 
We calculate $G^e$ and $G^{ph}$ as indicated in Eqs.~\ref{eq:land-e} and~\ref{eq:land-ph}, 
whereas $S$ and $\kappa^e$ are calculated from 
the electron transmission function (Eq.~\ref{eq:trans}) as 
follows~\cite{SivanPRB86,EsfarjaniPRB06},

\begin{equation}
S(\mu)= \frac{1}{e\,T}\frac{L_1(\mu)}{L_0(\mu)},
\label{eq_Seebeck}
\end{equation}

\begin{equation}
\kappa^e(\mu)=\frac{1}{T}\left\lbrace L_2(\mu) - \frac{L_1(\mu)^2}{L_0(\mu)} \right\rbrace,
 \label{eq_kappa_elec}
\end{equation}

\noindent where the functions $L_m(\mu)$ are defined as,

\begin{equation}
 L_m(\mu)=\frac{2}{h}\int_{-\infty}^{\infty} \mathcal{T}^e(E)(E-\mu)^m \left( -\frac{\partial f_0(E,\mu)}{\partial E} \right) dE.
 \label{eq_Lm}
\end{equation}

\section{Results}
\subsection{Electron transport}
\label{sec:elec-res}

We start our study by considering an armchair GNR with a full 
twist of $180\degree$ that develops over increasing lengths, $L_{twist}$.
Our goal is two-fold: (i)~estimating the minimum length required 
so that a $180\degree$ twist does not yield a significant decrease of the 
conductance (in the limit of an infinitely long scattering region 
the behaviour of the twisted system must approach that of the 
flat GNR); (ii)~characterizing the degradation of the conductance 
at those short lengths where the effect of the distortion is not negligible.

Some examples of the systems studied are sketched in Figure~\ref{fig:ri}.
We have analysed 12 different lengths of the twisted region,
ranging from $L_{twist} =9$ to 20 unit cells of the armchair GNR, i.e.
from 38.5 to 85~\AA. Notice that, prior to the calculation 
of the self-energies of the leads that allow treating the system 
as semi-infinite, the electronic structure within periodic boundary 
conditions must be calculated~\cite{BrandbygePRB02}. However, if the
GNR is symmetric with respect to the twist axis, a full twist of 
$180\degree$ naturally satisfies the requirement of periodicity. This is
not the case with the arbitrary twist angles that will be studied
in the second part of this section.
 
\begin{figure}[t]
\begin{center}
\includegraphics[width=0.8\columnwidth]{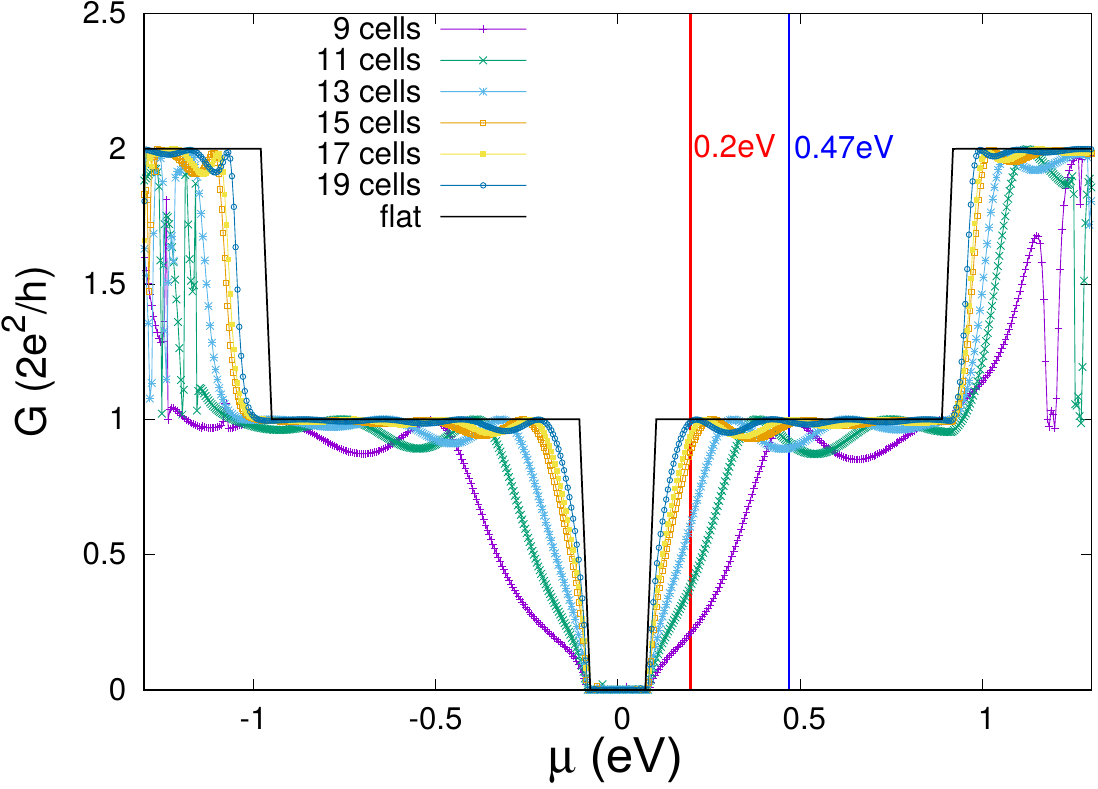}
\end{center}
\caption{Electrical conductance $G^e$ as a function of the chemical potential for a subset of the
         GNRs studied with a twist angle of 180\degree\ and different 
         length $L_{twist}$. The vertical lines indicate the energies
         at which we study the eigenchannels in Fig.~\ref{fig:wf}.}
\label{fig:Ts}
\end{figure}

The zero-bias conductance of a subset of 
the investigated systems is shown in Fig.~\ref{fig:Ts} together 
with that of the undistorted ribbon. We observe that the shorter is the length
within which the full twist is forced to occur, the lower
the conductance, with the most evident changes occurring for 
energy values next to the band edges and to the increase
from one to two transport channels ($\sim 1$~eV above (below)
the bottom (top) of the conduction (valence) band).
On the other hand, large values of the central length determine 
a generally high conductance, eventually approaching the value of 
the undistorted ribbon in the asymptotic case of infinite length.
As matter of fact, we obtain very similar conductances for 
lengths of the twisted regions larger than 14~unit cells
(see, e.g., the cases with $L_{twist}=$ 15, 17, and 19 in Fig.~\ref{fig:Ts}).

\begin{figure}[t]
\begin{center}
\includegraphics[width=0.8\textwidth]{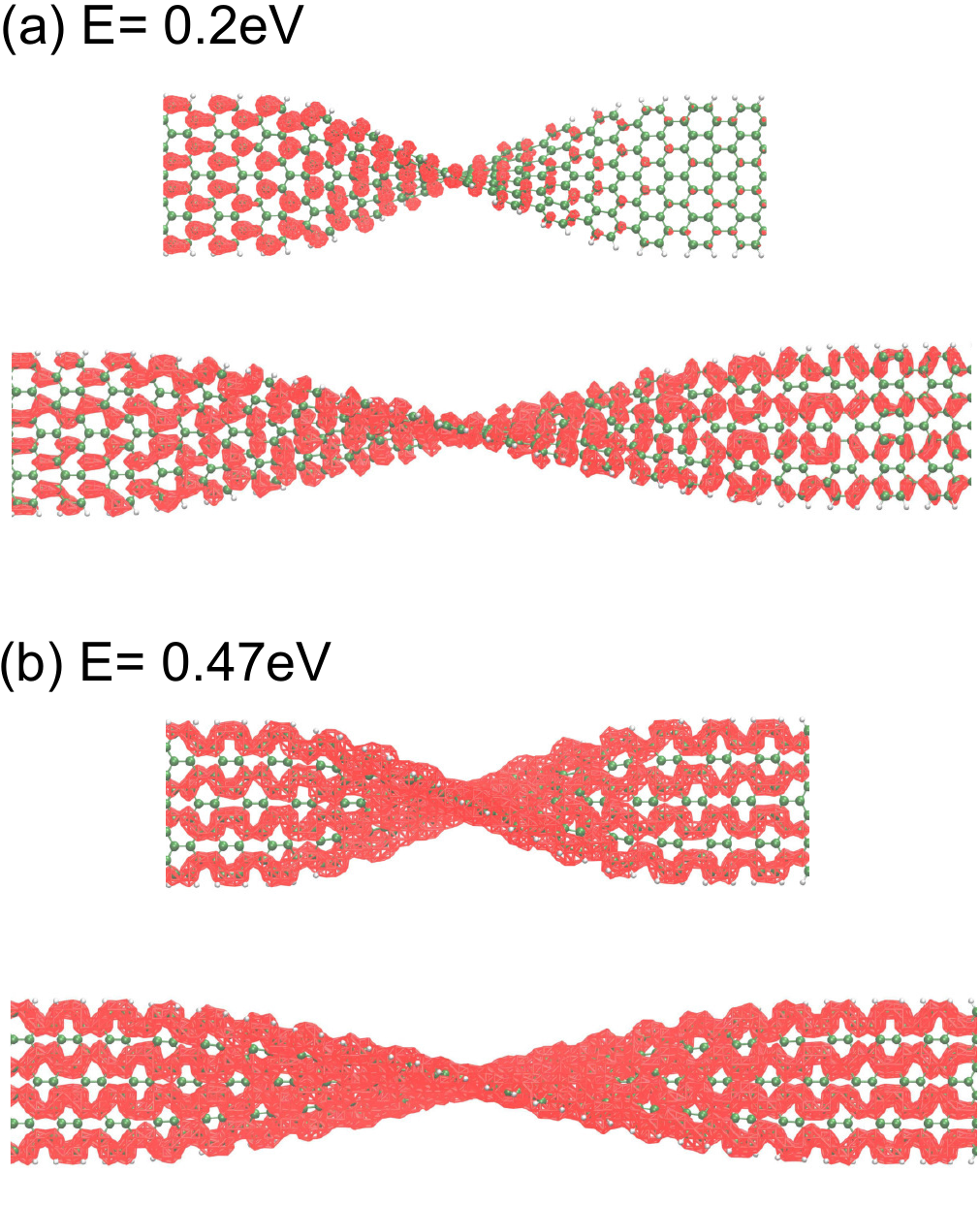}
\caption{Transport eigenchannels at (a)~$E=0.2$~eV and (b)~$E=0.47$~eV 
         for a short and a long GNR with a twist angle of 180\degree.
         The energies selected are indicated by vertical lines in
         Fig.~\ref{fig:Ts}.}
\label{fig:wf}
\end{center}
\end{figure}

Additional insight of the scattering mechanism is given by the analysis
of the transport eigenchannels. We have plotted the eigenchannel that carries
most of the current at two selected energies for a short and a long
GNR (Fig.~\ref{fig:wf}). At $E=0.2$~eV, the short GNR suffers a considerable scattering
and very little of the incoming eigenstate from the left-hand side
survives and is found on the right-hand side, after the $\pi$ twist.
For the longer GNR, on the other hand, the eigenstate is transmitted,
although after crossing the twist it changes its symmetry.
At $E=0.47$~eV the transmission is very high for all the GNR investigated,
as we already know from the analysis of the transmission. Accordingly,
the eigenchannel transmits well and similarly for both considered cases,
maintaining the same symmetry and a similar magnitude.

\begin{figure}[t]
\begin{center}
\includegraphics[width=0.8\columnwidth]{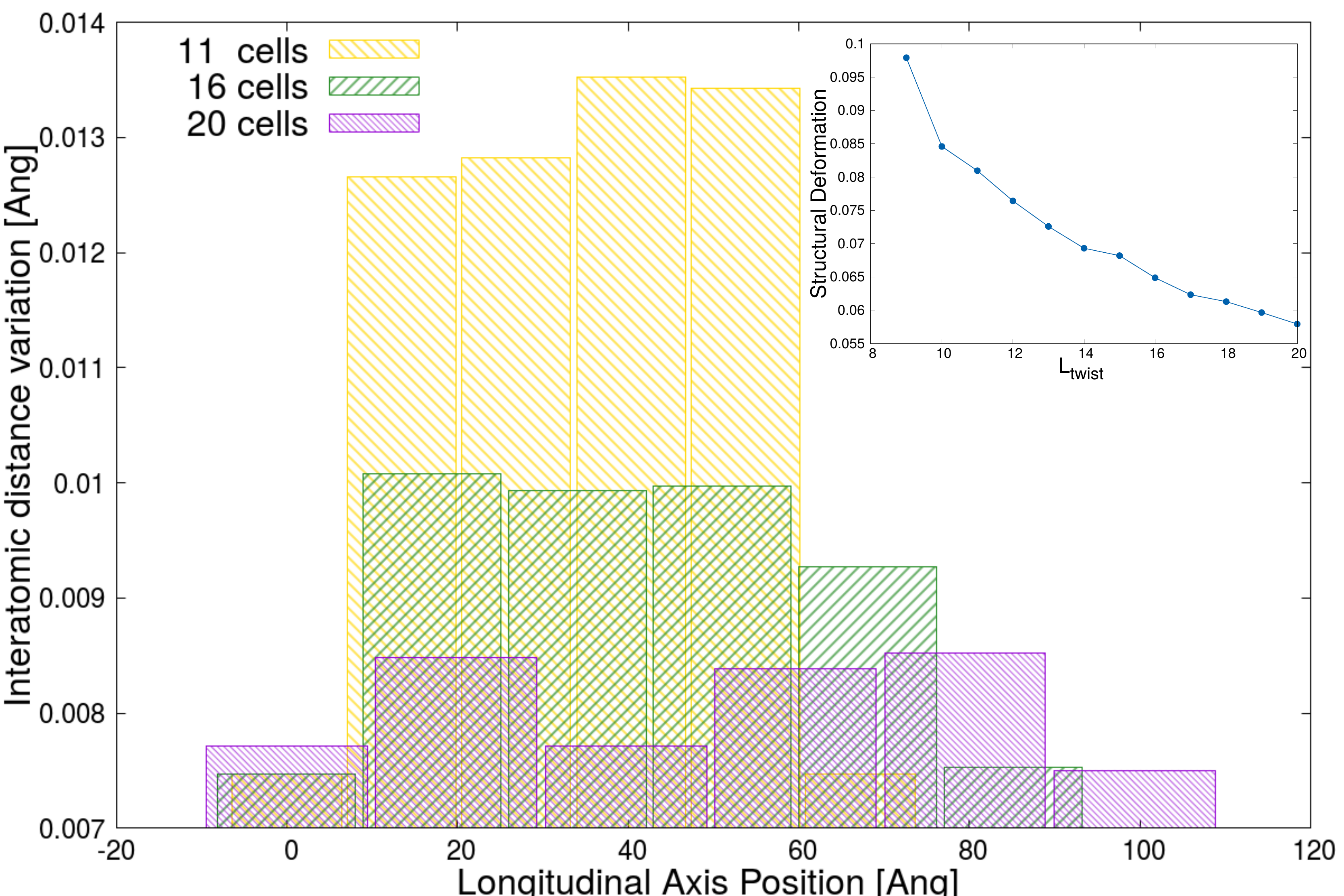}
\end{center}
\caption{Distribution of the average bond length variation as a function of
         the axial coordinate for a 11, a 16,
         and a 20 unit cell GNR with a twist angle of 180\degree.
         Inset: The standard deviation of the bond length as a function
         of $L_{twist}$; the curve asymptotically goes to zero for the 
         number of unit cells going to infinity.}
\label{fig:hist}
\end{figure}

In view of the results discussed above, we now give a closer look 
at what twisting a GNR means from the structural viewpoint. 
When a nanoribbon is twisted, it undergoes two main modifications
with respect to the planar configuration. On one hand, the structure 
is no longer flat; on the other side, the torsional distortion alters 
the bond-length distribution. Specifically, in an armchair GNR, twisting
affects mainly the components of the interatomic distances parallel to 
the transport direction, i.e. the ribbon axis. The extent to which 
the bond-lengths are modified is related to the length of the region 
over which the twisting takes place: the shorter the nanoribbon, the 
stronger the impact on the distances among the atoms. The dependence 
of this structural distortion on $L_{twist}$ is shown in fig.~\ref{fig:hist}, 
where the change in the bond length with respect to the 
flat ribbon is plotted as a function of the longitudinal position. 
The value is averaged over the number of atomic bonds in a given 
spatial interval. Each histogram corresponds to a specific length of 
twisted nanoribbon: a short, a medium and a long system ($L_{twist} = 11$, 
16 and 20, respectively).
The averaged deviation in the interatomic distance is almost uniform 
along the length of the twisted region (while it expectedly goes to 
zero at the extremes of the central scattering region where the geometry is kept frozen
during the relaxation to provide the correct coupling with the leads 
in the transport calculations). More importantly, 
however, Fig.~\ref{fig:hist} shows how the variation in interatomic distances 
tends to decrease for increasing lengths. A more concise representation
of these results is given in the inset, where by plotting 
the standard deviation of the bond length, we quantify 
the amount of local distortion as a function of $L_{twist}$.

\begin{figure}[t]
\begin{center}
\includegraphics[width=0.8\columnwidth]{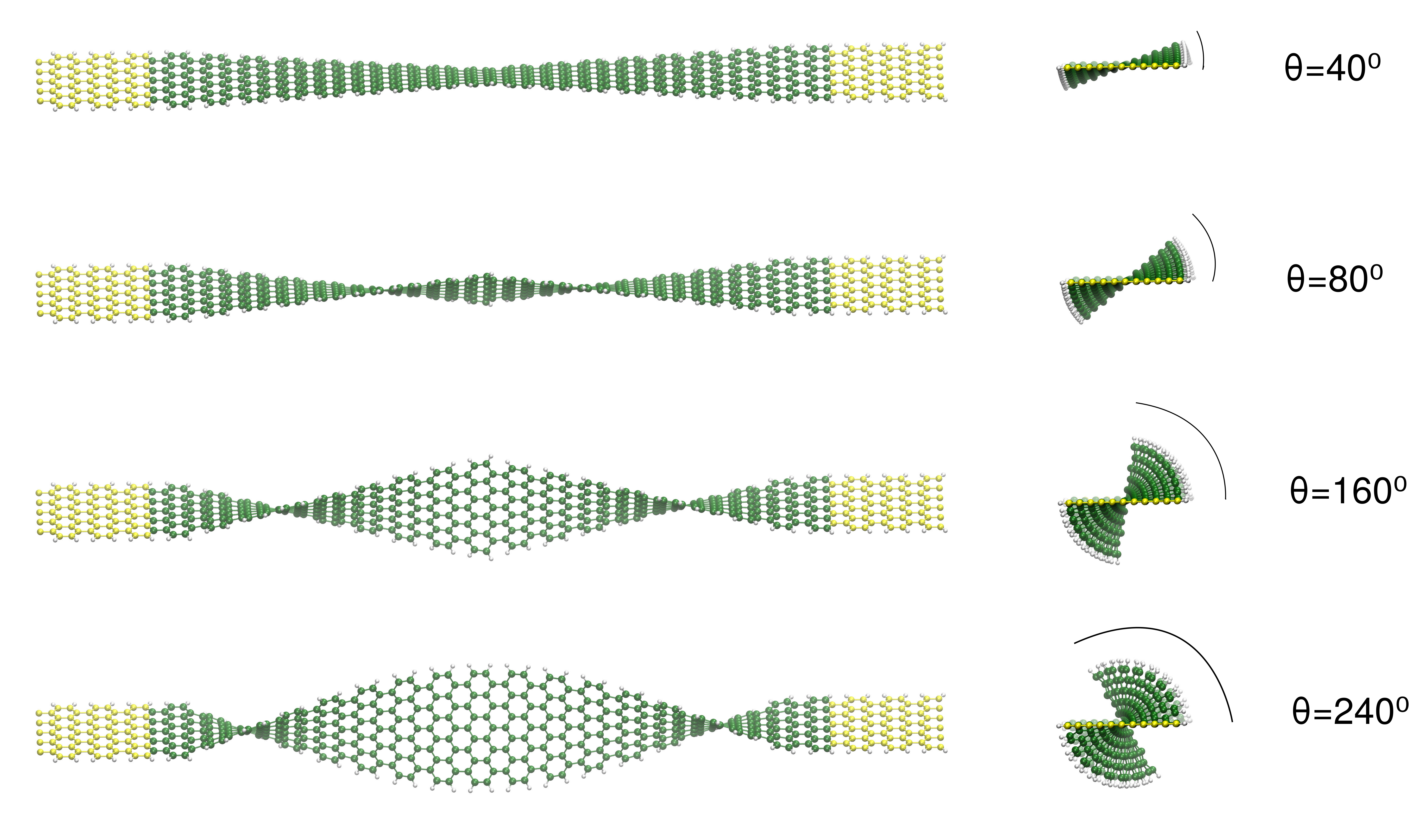}
\end{center}
\caption{Twisted armchair GNRs with a fixed length of the scattering
         region, $L_{twist}=16$ unit cells, and a twist angle $\phi$ of 
         (a)~20\degree, (b)~40\degree, (c)~800\degree, and 
         (d)~120\degree unit cells. Each of these rotations 
         develops over a length of $L_{twist}/2$ and is followed
         by a counter-rotation of the same magnitude, leading to a total
         rotation $\theta=2\phi$, to guarantee
         periodic boundary conditions. 
         }
\label{fig:part_twist}
\end{figure}

We now study the dependence of the conductance of an armchair GNR
in presence of a torsional deformation of an arbitrary angle.
As mentioned above, before coupling the scattering region to the
leads a calculation of its electronic
structure in periodic boundary conditions is needed. Notice that
such a requirement of periodicity of the twisted system is only 
guaranteed in the case of a torsion of $180 \degree$.
To avoid this limitation we study GNRs where a twist of $\phi$
that develops in the first half of the scattering region is
followed by an opposite twist of $-\phi$ in the remaining half.
Some atomistic structures of the studied systems are shown in 
Fig.~\ref{fig:part_twist}. 
Notice that, at variance with the case of a full twist, here we 
need to freeze a small region of the GNR, right where the
twist direction is reversed, to prevent the system to relax back
to the flat ground state.
We consider a fixed length of 16 unit cells of the scattering
region and vary the twist angle.

\begin{figure}[t]
\begin{center}
\includegraphics[width=0.8\columnwidth]{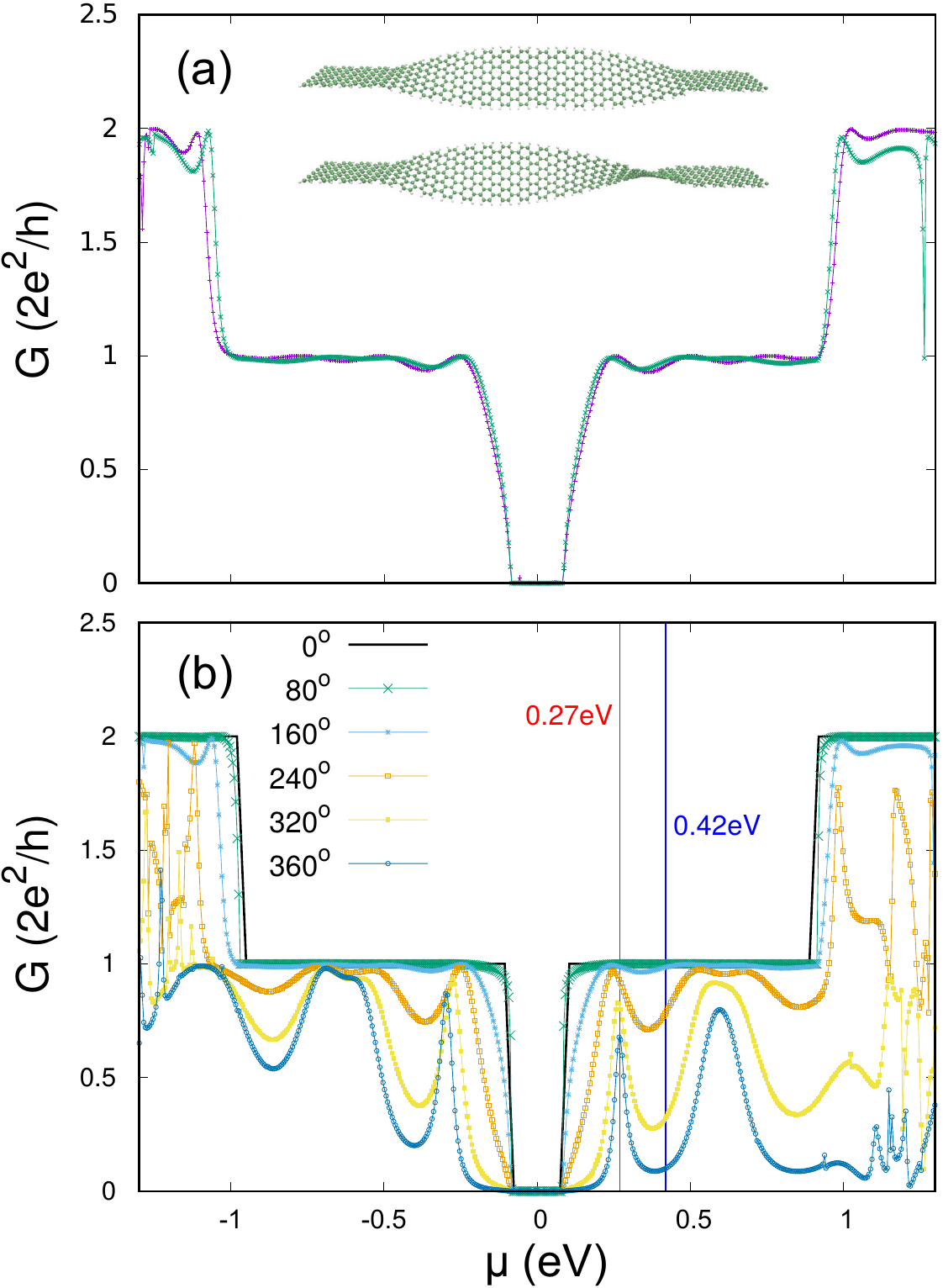}
\end{center}
\caption{(a) Comparison between the electrical conductance of a GNR with a twist
         of 90\degree\ followed by a counter-twist of $-90\degree$
         and a continuous twist of 180\degree. As the latter can be 
         seen as two twists of 90\degree\ one after the other, the
         only difference between the two cases is the reversal of 
         the twist angle at $L_{twist}/2$. The corresponding
         geometries are illustrated in the inset. 
         (b) Electrical conductance as a function of the chemical potential for 
         a subset of the GNRs studied with a length $L_{twist}$ of 16 unit cells 
         and different twist angles.
         }
\label{fig:trans_reversed_twist}
\end{figure}

Before performing a systematic study of the dependence on the angle,
we have analysed which is the effect of a reversal of the twist angle,
i.e. a torsion of $\phi$ followed by a torsion of $-\phi$. In
particular, we have considered the case of $\phi=90\degree$: a twist of 
$90\degree$ in the first half of the GNR, followed by a twist of 
$-90\degree$. We compare the transmission function of this case 
with a single, continuous twist of 180\degree\ that develops all over 
the length of the scattering region. The results are 
shown in Fig.~\ref{fig:trans_reversed_twist} (a). As can be seen there, the two
curves are almost indistinguishable. This means that reversing the 
twist angle has virtually no effect and a series of twists of  $+\phi/-\phi$ 
is roughly equivalent to a continuous twist of $2\phi$.
Therefore, we will present the results that follow as a function 
of $\theta=2\phi$, allowing direct comparison with the results 
discussed in the first part of this section.

Fig.~\ref{fig:trans_reversed_twist} (b) shows the conductance of the twisted GNR for 
different twisting angles $\theta$ in the range $[0,360\degree]$. 
As can be seen there, an increase of $\theta$ determines an increase 
of scattering in the central region with a following reduction in 
conductance. 
The curves at the highest twist angles ($\theta \geq 240\degree$) 
exhibit a strong chemical potential dependence, with a well recognizable peak 
structure that indicates that only at specific energies the transmission 
is finite, whereas it is almost fully suppressed elsewhere. 

\begin{figure}[t]
\begin{center}
\includegraphics[width=0.8\columnwidth]{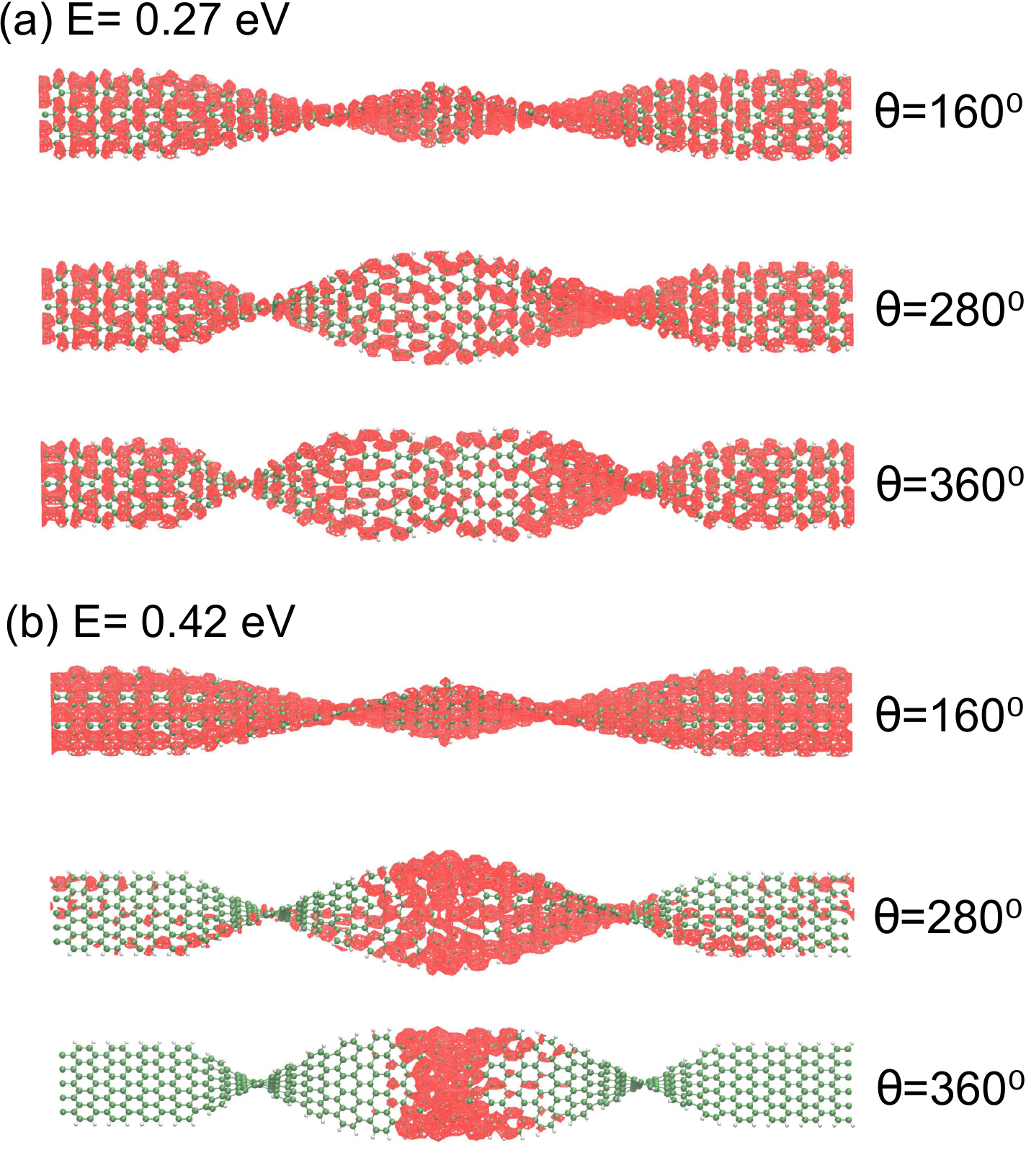}
\end{center}
\caption{Transport eigenchannels at (a)~$E=0.2$~eV and (b)~$E=0.47$~eV 
         for a GNR with twist angles $\theta=$ 160, 280 and 360\degree.
         The energies selected are indicated by vertical lines in
         Fig.~\ref{fig:trans_reversed_twist}(b).}
\label{fig:wfs}
\end{figure}

Therefore, it seems that for sufficiently large twist angles there
is a transition from a band-like one-dimensional conductance, to
a transport mechanism typical of molecular systems or constrictions, 
where energy is transmitted only in correspondence of resonant levels.
To better look into this transition, once again we study the 
transmission eigenchannels and their spatial distribution, which
we plot in Fig.~\ref{fig:wfs} at two selected energies. At $E=0.27$~eV,
where the transmission is high in all cases (see Fig.~\ref{fig:Ts})
no difference can be appreciated in the cases considered ($\theta=160\degree$,
280\degree, and 360\degree). The eigenchannels are mostly delocalized over 
the twisted region and guarantee a good coupling with the electrode incoming
states, thus resulting in a high transmission. 
On the other hand, moving to $E=0.42$~eV, where the conductance of GNR with 
high twist angles has a minimum, the eigenchannel is extended for small 
angles but becomes more and more localized as $\theta$ grows.
As it can be seen, there is an increasingly high probability of 
finding the electrons in the region between two {\em knots}, 
where the change in the rotation direction takes place.
This localized electronic states still allow electron transport,
but only at resonant energies with these {\em knot} levels.

\subsection{Phonon transport and thermoelectric figure of merit}

The effect of twisting a graphene nanoribbon on its phonon-mediated thermal 
transport properties has been previously studied with molecular dynamics~\cite{Chellattoan2013,Wang2014}
and NEGF simulations.~\cite{Wei2014} These studies have shown that, as for the 
case of electrical transport, the thermal conductance gets gradually reduced as 
a stronger torsional deformation is applied on the nanoribbon. In addition, the 
following trends have been observed: i) The conductance of zigzag terminated ribbons is 
more sensitive to twisting than that of armchair ones,~\cite{Wang2014} 
ii) the conductance is more effective reduced by twisting at low temperature regimes, 
when conductance degradation 
due to anharmonic phonon-phonon scattering is less relevant,~\cite{Chellattoan2013}
iii) twisting a nanoribbon has a sharper effect for narrower and shorter ribbons.~\cite{Wang2014,Wei2014}
In spite of these previous studies on the topic, it remains as an open question to 
determine the degree in which the torsional deformation affects the transport of 
specific phonon polarizations.

\begin{figure}[t]
\begin{center}
\includegraphics[width=0.8\columnwidth]{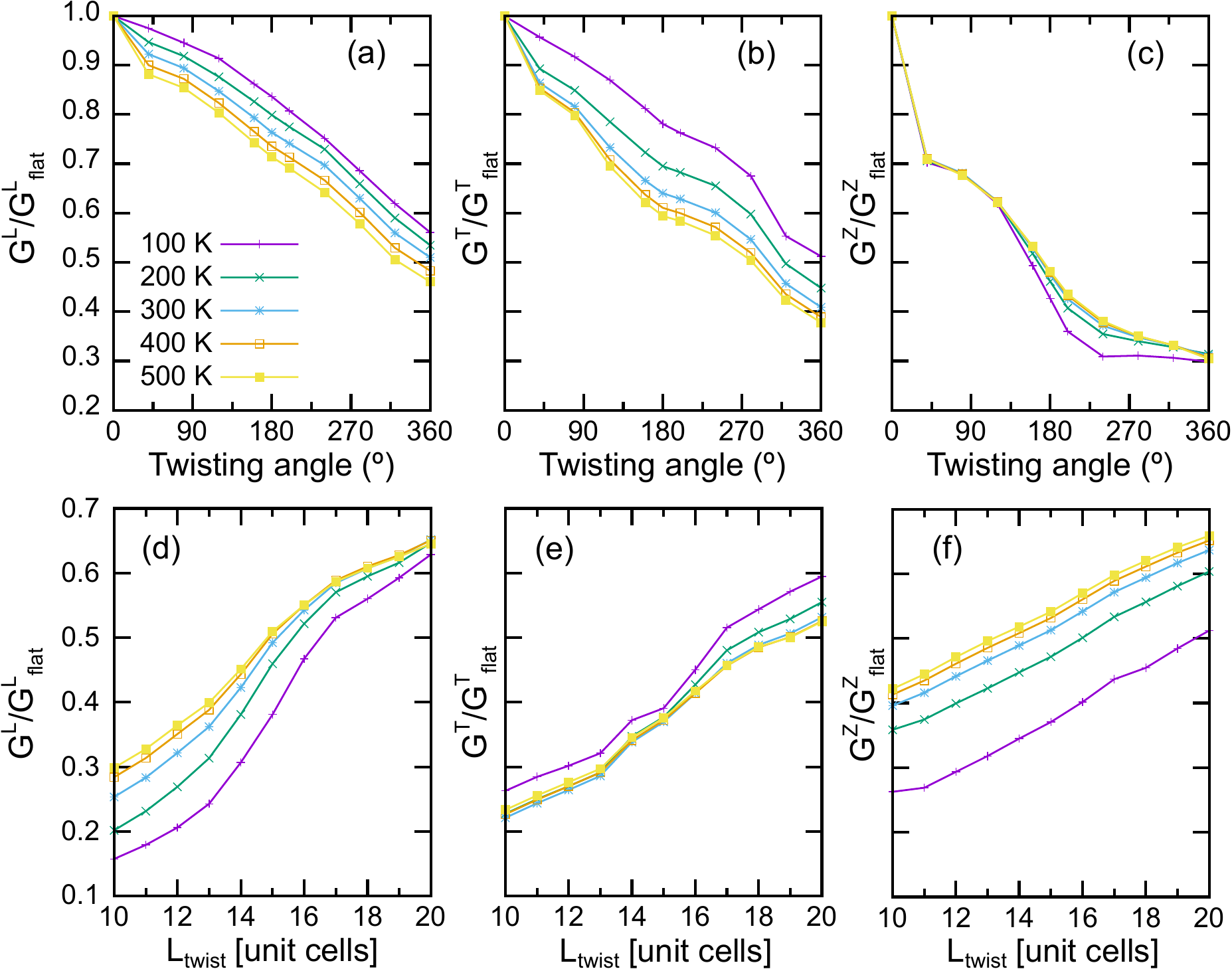}
\end{center}
\caption{Contribution to the thermal conductance from phonons polarized along the longitudinal (a,d), transverse (b,e) and
out-of-plane (c,f) directions scaled by their respective values for the flat nanoribbon. Upper panels show the results as a
function of the twisting angle for a fixed nanoribbon length. Bottom panels show the results as a function of the length of 
the twisted region for a fixed angle of 180$^{\circ}$. Different temperature results are shown as indicated.}
\label{fig:smt_phonon}
\end{figure}

Here we perform mode-dependent transport simulations and calculate the
contribution to the thermal conductance of twisted GNRs from longitudinal (L), transverse (T)
and out-of-plane (Z) phonons. To classify a phonon-mode as L, T or Z we 
adopt as a practical criterion that its eigenvector presents at least 70\% 
polarization along one of the three directions.
In Fig.~\ref{fig:smt_phonon} we show the results of the polarization-resolved
conductance analysis as the twisting angle is increased for a fixed nanoribbon
length of 16 unit cells (upper panels) and as the length of the twisted region is modified for
a fixed twisting angle of $180\degree$ (lower panels). The corresponding calculation on a flat 
nanoribbon yields that Z and L phonons contribute much more than T phonons 
to the thermal conductance (not shown). Upon twisting the ribbon we observe that
the conductance due to Z phonons experiences the largest degradation, of the order
of 70\% at 360$^{\circ}$, through a process that shows large independence on the 
temperature. This contrasts with the results of L and T phonons for which
twisting the ribbon reduces more effectively their contribution to the conductance 
at higher temperatures. On the other hand, upon shortening
the twisting region we observe that the conductance due to L phonons is the
most affected with respect to the flat nanoribbon case.

Finally we exploit the obtained results to assess the
performance of twisted GNR systems for thermoelectric applications. In this
field, a search for strategies to enhance charge transport and/or deteriorate
heat conductivity is currently ongoing for a variety of materials.
~\cite{RiffatATE03,DresselhausAdvMat07,ShakouriARMR11,ZebarjadiEES12,LiRMP12}
A measure for the performance of thermoelectric systems is given by the
figure of merit ZT (Eq.~\ref{eq_ZT}). In this respect, it is clear that the relative change of electron and
phonon conductance caused by twisting in our graphene nanoribbons can be
responsible for a large modification of thermoelectric figure of merit
with respect to the undistorted system. Indeed, depending on the relative
variation of the different ingredients of ZT, an improvement or
deterioration in the performance will follow. 

\begin{figure}[t]
\begin{center}
\includegraphics[width=0.8\columnwidth]{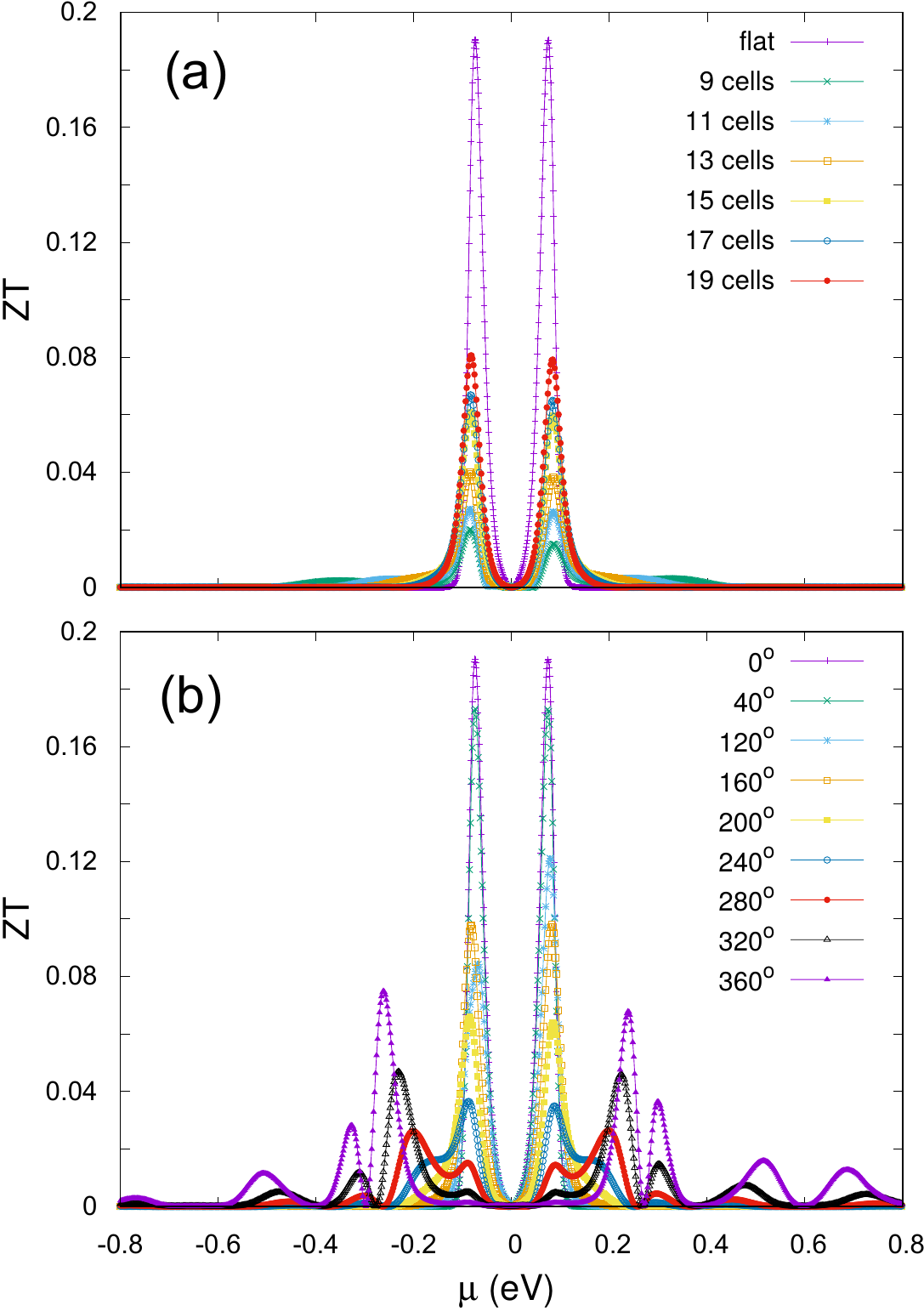}
\end{center}
\caption{Thermoelectric figure of merit as a function of the chemical potential
for GNRs with a fixed twist of 180$\degree $  and variable length (a) and for 
GNRs with a fixed length of 16 unit cells and different twisting angle (b).}
\label{fig:ZTvsmu}
\end{figure}

The evolution of ZT with the chemical potential at T=50 K is shown in Fig.~\ref{fig:ZTvsmu}
for different twisting lengths and angles.
Two general trends are observed. First, the chemical potential can be tuned in
order to increase the value of ZT, e.g. by applying a gate voltage to the
twisted channel. Second, the ZT is shown to have peaks in correspondence
of conduction and valence band edges.~\cite{ZhengAPL12}
As can be seen in Fig.~\ref{fig:ZTvsmu} (a), the ZT peaks decrease when the twisting 
length is reduced and their value is always smaller than that of the undistorted ribbon.
The figure of merit is expected to tend asymptotically to this maximum
value for sufficiently large twisting lengths.
Regarding the effect of the twist angle (Fig.~\ref{fig:ZTvsmu} (b)),
we observe that ZT peaks are also found at energies
close to the band edges for sufficiently small angles.
However, when the twisting is made stronger ($> 280\degree$), a new
collection of peaks appears at new values of chemical potential corresponding to 
maxima of electronic conductance (see Fig.~\ref{fig:Ts}).
This behaviour consistently reflects the discussed change in the nature of
electron transport in these systems, evolving from 1D semiconducting
channels to  molecule-like systems. Notwithstanding, it is important to note that, 
also for this differently twisted ribbons, the figure of merit is smaller than in the 
untwisted case. We have checked that this general observation is ruled by the
reduction of the Seebeck coefficient as the GNR is deformed.

\section{Conclusions}
\label{sec:concl}

In this work we simulated ballistic electron and phonon transport 
in graphene nanoribbons with torsional deformations through a
nonequilibrium Green's functions procedure and determined the following.
Broadly speaking, both electrical and thermal conductances are reduced by 
twisting a nanoribbon. The shorter is the length within which a
given twist is applied, the larger is the structural deformation,
which is reflected in the distribution of atomic bonds lengths, and
the lower is the electrical conductance. In turn, deformations applied
over a long portion of nanoribbon have a negligible effect on the 
transport properties. On the other hand, when two consecutive twists are applied
in opposite angular directions a transition from a 1D to a 0D (molecular)
electron transport regime is observed. The molecular-like transport
is a consequence of the formation of localized states in the region
of the nanoribbon between the two knots. Concerning phonon transport, 
we have observed that the propagation of out-of-plane and longitudinal vibrational modes is the most impeded 
by increasing the twist angle and by shortening the twist length, respectively.
Finally, the thermoelectric efficiency of GNRs is generally attenuated 
by torsional deformations because these entail a reduction in
the Seebeck coefficient, thereby reducing the figure-of-merit ZT.

\section*{Acknowledgements}
We thank Troels Markussen for useful discussions.
M.R. and R.R. acknowledge financial support by the Ministerio de Econom\'ia y
Competitividad (MINECO) under grant FEDER-MAT2013-40581-P and by the
Severo Ochoa Centres of Excellence Program under Grant SEV-2015-0496
and by the Generalitat de Catalunya under grants no. 2014 SGR 301 and
through the Beatriu de Pin\'os fellowship program (2014 BP\_B 00101).
We thank the Centro de Supercomputaci\'on de Galicia (CESGA) for the 
use of their computational resources.

\section*{References}
\bibliography{twist}

\providecommand{\newblock}{}
\begin{thebibliography}{10}
\expandafter\ifx\csname url\endcsname\relax
  \def\url#1{{\tt #1}}\fi
\expandafter\ifx\csname urlprefix\endcsname\relax\def\urlprefix{URL }\fi
\providecommand{\eprint}[2][]{\url{#2}}

\bibitem{novoselov2004electric}
Novoselov K~S, Geim A~K, Morozov S~V, Jiang D, Zhang Y, Dubonos S~V, Grigorieva
  I~V and Firsov A~A 2004 {\em Science\/} {\bf 306} 666--669

\bibitem{novoselov2005two}
Novoselov K~S, Geim A~K, Morozov S, Jiang D, Katsnelson M, Grigorieva I,
  Dubonos S and Firsov A 2005 {\em Nature\/} {\bf 438} 197--200

\bibitem{zhang2005experimental}
Zhang Y, Tan Y~W, Stormer H~L and Kim P 2005 {\em Nature\/} {\bf 438} 201--204

\bibitem{geim2007rise}
Geim A~K and Novoselov K~S 2007 {\em Nat. Mater.\/} {\bf 6} 183--191

\bibitem{neto2009electronic}
Neto A~C, Guinea F, Peres N~M, Novoselov K~S and Geim A~K 2009 {\em Rev. Mod.
  Phys.\/} {\bf 81} 109

\bibitem{novoselov2011nobel}
Novoselov K 2011 {\em Rev. Mod. Phys.\/} {\bf 83} 837

\bibitem{geim2011nobel}
Geim A~K 2011 {\em Rev. Mod. Phys.\/} {\bf 83} 851

\bibitem{ButlerACSNano13}
Butler S~Z, Hollen S~M, Cao L, Cui Y, Gupta J~A, Guti{\'e}rrez H~R, Heinz T~F,
  Hong S~S, Huang J, Ismach A~F, Johnston-Halperin E, Kuno M, Plashnitsa V~V,
  Robinson R~D, Ruoff R~S, Salahuddin S, Shan J, Shi L, Spencer M~G, Terrones
  M, Windl W and Goldberger J~E 2013 {\em ACS Nano\/} {\bf 7} 2898--2926

\bibitem{BhimanapatiACSNano15}
Bhimanapati G~R, Lin Z, Meunier V, Jung Y, Cha J, Das S, Xiao D, Son Y, Strano
  M~S, Cooper V~R, Liang L, Louie S~G, Ringe E, Zhou W, Kim S~S, Naik R~R,
  Sumpter B~G, Terrones H, Xia F, Wang Y, Zhu J, Akinwande D, Alem N, Schuller
  J~A, Schaak R~E, Terrones M and Robinson J~A 2015 {\em ACS Nano\/} {\bf 9}
  11509--11539

\bibitem{NovoselovScience16}
Novoselov K~S, Mishchenko A, Carvalho A and {Castro Neto} A~H 2016 {\em
  Science\/} {\bf 353}

\bibitem{gunlycke2007room}
Gunlycke D, Lawler H and White C 2007 {\em Phys. Rev. B\/} {\bf 75} 085418

\bibitem{novoselov2007room}
Novoselov K~S, Jiang Z, Zhang Y, Morozov S, Stormer H~L, Zeitler U, Maan J,
  Boebinger G, Kim P and Geim A~K 2007 {\em Science\/} {\bf 315} 1379--1379

\bibitem{balandin2011thermal}
Balandin A~A 2011 {\em Nat. Mater.\/} {\bf 10} 569--581

\bibitem{du2008approaching}
Du X, Skachko I, Barker A and Andrei E~Y 2008 {\em Nat. Nanotechnol.\/} {\bf 3}
  491--495

\bibitem{bonaccorso2010graphene}
Bonaccorso F, Sun Z, Hasan T and Ferrari A 2010 {\em Nat. Photonics\/} {\bf 4}
  611--622

\bibitem{bonaccorso2015graphene}
Bonaccorso F, Colombo L, Yu G, Stoller M, Tozzini V, Ferrari A~C, Ruoff R~S and
  Pellegrini V 2015 {\em Science\/} {\bf 347} 1246501

\bibitem{li2008chemically}
Li X, Wang X, Zhang L, Lee S and Dai H 2008 {\em Science\/} {\bf 319}
  1229--1232

\bibitem{jiao2009narrow}
Jiao L, Zhang L, Wang X, Diankov G and Dai H 2009 {\em Nature\/} {\bf 458}
  877--880

\bibitem{cai2010atomically}
Cai J, Ruffieux P, Jaafar R, Bieri M, Braun T, Blankenburg S, Muoth M,
  Seitsonen A~P, Saleh M, Feng X {\em et~al.\/} 2010 {\em Nature\/} {\bf 466}
  470--473

\bibitem{hashimoto2004direct}
Hashimoto A, Suenaga K, Gloter A, Urita K and Iijima S 2004 {\em Nature\/} {\bf
  430} 870--873

\bibitem{gass2008free}
Gass M~H, Bangert U, Bleloch A~L, Wang P, Nair R~R and Geim A {\em Nat.
  Nanotechnol.\/} {\bf 3} 676

\bibitem{meyer2008direct}
Meyer J~C, Kisielowski C, Erni R, Rossell M~D, Crommie M and Zettl A 2008 {\em
  Nano Lett.\/} {\bf 8} 3582--3586

\bibitem{girit2009graphene}
Girit {\c{C}}~{\"O}, Meyer J~C, Erni R, Rossell M~D, Kisielowski C, Yang L,
  Park C~H, Crommie M, Cohen M~L, Louie S~G {\em et~al.\/} 2009 {\em Science\/}
  {\bf 323} 1705--1708

\bibitem{warner2009structural}
Warner J~H, R{\"u}mmeli M~H, Ge L, Gemming T, Montanari B, Harrison N~M,
  B{\"u}chner B and Briggs G~A~D 2009 {\em Nat. Nanotechnol.\/} {\bf 4}
  500--504

\bibitem{meyer2007structure}
Meyer J~C, Geim A~K, Katsnelson M~I, Novoselov K~S, Booth T~J and Roth S 2007
  {\em Nature\/} {\bf 446} 60--63

\bibitem{brivio2011ripples}
Brivio J, Alexander D~T and Kis A 2011 {\em Nano Lett.\/} {\bf 11} 5148--5153

\bibitem{kittel1966introduction}
Kittel C 1966 {\em {Introduction to solid state}\/} vol 162 (John Wiley \&
  Sons)

\bibitem{banhart2010structural}
Banhart F, Kotakoski J and Krasheninnikov A~V 2010 {\em ACS Nano\/} {\bf 5}
  26--41

\bibitem{liu2015defects}
Liu L, Qing M, Wang Y and Chen S 2015 {\em J. Mater. Sci. Technol.\/} {\bf 31}
  599--606

\bibitem{haskins2011control}
Haskins J, K{\i}nac{\i} A, Sevik C, Sevin{\c{c}}li H, Cuniberti G and
  {\c{C}}ag{\i}n T 2011 {\em ACS Nano\/} {\bf 5} 3779--3787

\bibitem{fasolino2007intrinsic}
Fasolino A, Los J and Katsnelson M~I 2007 {\em Nat. Mater.\/} {\bf 6} 858--861

\bibitem{cresti2008charge}
Cresti A, Nemec N, Biel B, Niebler G, Triozon F, Cuniberti G and Roche S 2008
  {\em Nano Res.\/} {\bf 1} 361--394

\bibitem{sevinccli2010enhanced}
Sevin{\c{c}}li H and Cuniberti G 2010 {\em Phys. Rev. B\/} {\bf 81} 113401

\bibitem{li2011efficient}
Li W, Sevin{\c{c}}li H, Roche S and Cuniberti G 2011 {\em Phys. Rev. B\/} {\bf
  83} 155416

\bibitem{li2010phonon}
Li W, Sevin{\c{c}}li H, Cuniberti G and Roche S 2010 {\em Phys. Rev. B\/} {\bf
  82} 041410

\bibitem{areshkin2007ballistic}
Areshkin D~A, Gunlycke D and White C~T 2007 {\em Nano Lett.\/} {\bf 7} 204--210

\bibitem{bellido2012graphene}
Bellido E~P and Seminario J~M 2012 {\em J. Phys. Chem. C\/} {\bf 116}
  8409--8416

\bibitem{cadelano2010interplay}
Cadelano E, Giordano S and Colombo L 2010 {\em Phys. Rev. B\/} {\bf 81} 144105

\bibitem{koskinen2012graphene}
Koskinen P 2012 {\em Phys. Rev. B\/} {\bf 85} 205429

\bibitem{sanders2013theory}
Sanders G, Nugraha A, Sato K, Kim J, Kono J, Saito R and Stanton C 2013 {\em J.
  Phys.: Condens. Matter\/} {\bf 25} 144201

\bibitem{lopez2011nanostructured}
L{\'o}pez-Su{\'a}rez M, Rurali R, Gammaitoni L and Abadal G 2011 {\em Phys.
  Rev. B\/} {\bf 84} 161401

\bibitem{lopez2013buckling}
L{\'o}pez-Su{\'a}rez M, Rurali R and Abadal G 2013 {\em Microelectron. Eng.\/}
  {\bf 111} 122--125

\bibitem{lopez2014piezoelectric}
L{\'o}pez-Su{\'a}rez M, Pruneda M, Abadal G and Rurali R 2014 {\em
  Nanotechnology\/} {\bf 25} 175401

\bibitem{KhlobystovACSNANO11}
Khlobystov A~N 2011 {\em ACS Nano\/} {\bf 5} 9306--9312

\bibitem{ChamberlainACSNANO12}
Chamberlain T~W, Biskupek J, Rance G~A, Chuvilin A, Alexander T~J, Bichoutskaia
  E, Kaiser U and Khlobystov A~N 2012 {\em ACS Nano\/} {\bf 6} 3943--3953

\bibitem{EliasNL10}
El{\'i}as A~L, Botello-M{\'e}ndez A~R, Meneses-Rodr{\'i}guez D, {Jehov{\'a}
  Gonz{\'a}lez} V, Ram{\'i}rez-Gonz{\'a}lez D, Ci L, Mu{\~n}oz-Sandoval E,
  Ajayan P~M, Terrones H and Terrones M 2010 {\em Nano Letters\/} {\bf 10}
  366--372

\bibitem{Cranford2011}
Cranford S and Buehler M~J 2011 {\em Model. Simul. Mater. Sci. Eng.\/} {\bf 19}
  54003

\bibitem{Nikiforov2014}
Nikiforov I, Hourahine B, Frauenheim T and Dumitric\u{a} T 2014 {\em J. Phys.
  Chem. Lett.\/} {\bf 5} 4083--4087

\bibitem{Sadrzadeh2011}
Sadrzadeh A, Hua M and Yakobson B~I 2011 {\em Appl. Phys. Lett.\/} {\bf 99}
  13102

\bibitem{Al-Aqtash2013}
Al-Aqtash N, Li H, Wang L, Mei W~N and Sabirianov R 2013 {\em Carbon N. Y.\/}
  {\bf 51} 102--109 ISSN 00086223

\bibitem{Jia2014}
Jia J, Shi D, Feng X and Chen G 2014 {\em Carbon N. Y.\/} {\bf 76} 54--63 ISSN
  00086223

\bibitem{Xu2015}
Xu N, Huang B, Li J and Wang B 2015 {\em Solid State Commun.\/} {\bf 202}
  39--42 ISSN 00381098

\bibitem{Tang2012}
Tang G~P, Zhou J~C, Zhang Z~H, Deng X~Q and Fan Z~Q 2012 {\em Appl. Phys.
  Lett.\/} {\bf 101}

\bibitem{SolerJPCM02}
Soler J~M, Artacho E, Gale J~D, Garc{\'i}a A, Junquera J, Ordej{\'o}n P and
  S{\'a}nchez-Portal D 2002 {\em J. Phys.: Condens. Matter\/} {\bf 14}
  2745--2779

\bibitem{BrandbygePRB02}
Brandbyge M, Mozos J~L, Ordej{\'o}n P, Taylor J and Stokbro K 2002 {\em Phys.
  Rev. B\/} {\bf 65} 165401

\bibitem{BrennerPRB90}
Brenner D~W 1990 {\em Phys. Rev. B\/} {\bf 42}(15) 9458--9471

\bibitem{GaleMS03}
Gale J~D and Rohl A~L 2003 {\em Mol. Simulat.\/} {\bf 29} 291--341

\bibitem{Ong2015}
Ong Z~Y and Zhang G 2015 {\em Phys. Rev. B\/} {\bf 91} 174302 ISSN 1550235X

\bibitem{SivanPRB86}
Sivan U and Imry Y 1986 {\em Phys. Rev. B\/} {\bf 33}(1) 551--558

\bibitem{EsfarjaniPRB06}
Esfarjani K, Zebarjadi M and Kawazoe Y 2006 {\em Phys. Rev. B\/} {\bf 73}(8)
  085406

\bibitem{Chellattoan2013}
Chellattoan R and Sathian S~P 2013 {\em Solid State Commun.\/} {\bf 173} 1--4
  ISSN 00381098

\bibitem{Wang2014}
Wang F, Drzal L~T, Qin Y and Huang Z 2014 {\em J. Mater. Sci.\/} {\bf 50}
  1082--1093 ISSN 0022-2461

\bibitem{Wei2014}
Wei X, Guo G, Tao O and Xiao H 2014 {\em J. Appl. Phys.\/} {\bf 115} 2012--2017
  ISSN 10897550

\bibitem{RiffatATE03}
Riffat S and Ma X 2003 {\em Appl. Therm. Eng.\/} {\bf 23} 913--935

\bibitem{DresselhausAdvMat07}
Dresselhaus M~S, Chen G, Tang M~Y, Yang R~G, Lee H, Wang D~Z, Ren Z~F, Fleurial
  J~P and Gogna P 2007 {\em Adv. Mater.\/} {\bf 19} 1043--1053

\bibitem{ShakouriARMR11}
Shakouri A 2011 {\em Annu. Rev. Mater. Res.\/} {\bf 41} 399--431

\bibitem{ZebarjadiEES12}
Zebarjadi M, Esfarjani K, Dresselhaus M~S, Ren Z~F and Chen G 2012 {\em Energy
  Environ. Sci.\/} {\bf 5}(1) 5147--5162

\bibitem{LiRMP12}
Li N, Ren J, Wang L, Zhang G, H{\"a}nggi P and Li B 2012 {\em Rev. Mod.
  Phys.\/} {\bf 84}(3) 1045--1066

\bibitem{ZhengAPL12}
Zheng H, Liu H~J, Tan X~J, Lv H~Y, Pan L, Shi J and Tang X~F 2012 {\em Appl.
  Phys. Lett.\/} {\bf 100} ISSN 00036951

\end{thebibliography}

\clearpage

\clearpage

\clearpage

\clearpage

\clearpage

\clearpage

\clearpage

\end{document}